\documentclass[conference]{IEEEtran}

\newif\iffull

\usepackage{algorithmic, algorithm}
\usepackage[english]{babel}
\usepackage{amssymb}
\usepackage{amsmath}
\usepackage[dvipsnames]{xcolor}
\usepackage{cite}
\usepackage{graphicx}
\usepackage{balance}
\usepackage{amsthm}
\usepackage[hidelinks]{hyperref}

\theoremstyle{definition}
\newtheorem{theorem}{Theorem}
\newtheorem{lemma}{Lemma}

\newtheorem{construction}{Construction}
\newtheorem{corollary}{Corollary}
\newtheorem{definition}{Definition}

\newtheorem{example}{Example}

\newtheorem{problem}{Problem}
\newtheorem{claim}{Claim}

\usepackage{hyperref}
\addto\extrasenglish{

}

\newcommand{\cC}{{\cal C}}

\newcommand{\cN}{{\cal N}}

\newcommand{\cP}{{\cal P}}

\newcommand{\cT}{{\cal T}}

\newcommand{\cV}{{\cal V}}

\newcommand{\bfw}{{\boldsymbol w}}

\newcommand{\mA}{{\mathsf A}}
\newcommand{\mC}{{\mathsf C}}
\newcommand{\mG}{{\mathsf G}}
\newcommand{\mT}{{\mathsf T}}

\def\CC#1#2{\ensuremath{\left(\kern-.3em\left(\genfrac{}{}{0pt}{}{#1}{#2}\right)\kern-.3em\right)}}

\title{Representing Information on DNA using Patterns Induced by Enzymatic Labeling\vspace{-0.8ex}}
\author{\IEEEauthorblockN{Daniella Bar-Lev\IEEEauthorrefmark{1}, Tuvi Etzion\IEEEauthorrefmark{1}, Eitan Yaakobi\IEEEauthorrefmark{1}, and Zohar Yakhini\IEEEauthorrefmark{1}\IEEEauthorrefmark{2}}
\thanks{The work of D. Bar-Lev, E. Yaakobi, and Z. Yakhini was funded by the European Union (DiDAX, 101115134). Views and opinions expressed are however those of the author(s) only and do not necessarily reflect those of the European Union or the European Research Council Executive Agency. Neither the European Union nor the granting authority can be held responsible for them. The work of D. Bar-Lev and T. Etzion was supported in part by the Israel Science Foundation (ISF) under Grant 222/19 and in part by the Technion Data Science Initiative.}
\IEEEauthorblockA{\IEEEauthorrefmark{1}\textit{Faculty of Computer Science, Technion -- Israel Institute of Technology, Haifa, Israel} \\
\IEEEauthorblockA{\IEEEauthorrefmark{2}\textit{School of Computer Science, Reichman University, Herzliya, Israel}}
Email: \{daniellalev, yaakobi, etzion, zohary\}@cs.technion.ac.il}}

\IEEEoverridecommandlockouts
\IEEEaftertitletext{\vspace{-5pt}}

\begin{document}

\maketitle
\vspace{-2ex}
\begin{abstract}
Enzymatic DNA labeling is a powerful tool 
with applications in biochemistry, molecular biology, biotechnology, medical science, and genomic research. This paper contributes to the evolving field of DNA-based data storage by presenting a formal framework for modeling DNA labeling in strings, specifically tailored for data storage purposes. Our approach involves a known DNA molecule as a template for labeling, employing patterns induced by a set of designed labels to represent information. 
One hypothetical implementation can use CRISPR-Cas9 and gRNA reagents for labeling.
Various aspects of the general labeling channel, including fixed-length labels, are explored, and upper bounds on the maximal size of the corresponding codes are given. The study includes the development of an efficient encoder-decoder pair that is proven optimal in terms of maximum code size under specific conditions.
\end{abstract}

\vspace{-1.9ex}
\section{Introduction}
\label{sec:introduction}
Enzymatic DNA labeling is a powerful tool 
with applications in biochemistry, molecular biology, biotechnology, medical science, and genomic research~\cite{
moter2000fluorescence, chen2018efficient, gruszka2021single,
ma2015multicolor,
deen2017methyltransferase,
young2020technical}. The technique involves the deliberate labeling of specific parts of the DNA molecule with specific markers. In recent literature, it was also demonstrated that DNA labeling (and more generally, DNA editing) can be used in the emerging technology of data storage on DNA. see e.g., \cite{tabatabaei2020dna, Sadremomtaz2023storageCRISPR}. One important approach to labeling is the use of CRISPR-Cas9 based systems \cite{jinek2012programmable, chen2018efficient, cong2013multiplex, amit2021crispector}. 

Beyond its practical applications, recent studies shifted focus towards understanding DNA labeling from an information theoretic perspective. In~\cite{hanania2023capacity} Hanania et. al. modeled the process of DNA labeling as a communication channel. In this setup, either a fixed single label or a small set of fixed labels is being used. The input is a 4-ary sequence representing the DNA molecule and the output is a sequence in which \emph{non-zero} entries represent the presence of labels in the molecule. The work presents multiple results regarding the capacity of this channel, as a function of label(s) used. Another related communication channel, which was studied by Nogin et. al.~\cite{Nogin2023OM}, considered the process of Optical Genome Mapping (OGM)~\cite{levy2013beyond, muller2017optical}, a useful application of DNA labeling that involves optically imaging DNA fragments containing labeled short sequence patterns. The study utilizes techniques from information theory to enable the design of optimal labeling patterns for this application. In \cite{tabatabaei2020dna} the authors introduced and investigated concept related to native DNA to store information using detectable chemical modifications. 

 In this work, we define a formal framework for modeling labeling in DNA strings and present results pertaining to optimal design in such systems, for the purpose of data storage in DNA. 
 In contrast to the model suggested in~\cite{hanania2023capacity}, in our model a specific known DNA molecule is considered as template for labeling, while the information is represented by patterns induced by a set of labels designed for this purpose. More specifically, we code any given message to a set of   labels that will induce a unique pattern. When reading the information, post communication, the output of the channel is a binary vector representing the sites labeled by the selected labels in the known  template DNA molecule. 
 
 We characterize the performance of this labeling channel addressing several possible variants. In particular, we consider labels of fixed length $\ell$, which is the expected case if CRISPR-Cas9 were to be used, as well as systems that can support a dynamic range of lengths. We developed, for example, an efficient encoder and decoder pair for this channel in the special case in which $\ell$ is fixed and the template sequence is $\ell$-\emph{repeat-free}. In this case, we proved optimality of our construction. 

The rest of the paper is organized as follows. In \autoref{sec:background}, we introduce the definitions that will be utilized throughout the paper, including the formal definitions of our labeling model and the problems investigated. Additionally, \autoref{sec:background} provides several bounds that consider strings with a small period as templates.
Moving on to \autoref{sec: fix length}, our focus shifts to labels with a fixed length, denoted as $\ell$. This section presents both upper and lower bounds on the maximum size of codes in our model. Furthermore, we introduce an efficient construction that achieves the upper bound.

\vspace{-0.5ex}
\section{Definitions, Problem Statement, and a First Bound}
\label{sec:background}
\vspace{-0.5ex}
\subsection{Definitions}\label{subsec:Definitions}
Let $\Sigma = \{\mathsf{A,C,G,T}\}$ denote the DNA alphabet, and let $\Sigma^n$ denote the set of all length-$n$ sequences over $\Sigma$ and $\Sigma^*$ is the set of all sequences of any length over $\Sigma$. Denote by $[n]$ the set $\{1, 2,\ldots, n\}$.
For a sequence $S=(s_1,\ldots,s_n)\in\Sigma^n$, and $1\leq i \leq n-\ell+1$ , let $S_{[i;\ell]}\triangleq(s_i,\ldots, s_{i+\ell-1})$ and let  $W_\ell(S)\triangleq \{S_{[i;\ell]}: 1\le i\le n-\ell+1\}$. A \emph{label} $\lambda\in\Sigma^*$ is a relatively short sequence over $\Sigma$. For a sequence $S\in\Sigma^n$, the \emph{labeling} of $S$ with the label $\lambda$ is defined as follows.
\begin{definition}
Given a label $\lambda\in\Sigma^\ell$, the \emph{$\lambda$-labeling function}, $\lambda:\Sigma^*\to\{0,1\}^*$, is a mapping that corresponds to the label $\lambda$, where the \emph{$\lambda$-labeling} of a sequence $S\in\Sigma^n$ is defined as a binary sequence $\lambda(S)\in \{0,1\}^n$ in which $\lambda(S)_{[i:\ell]}=(1, 1, \ldots, 1)$ if and only if $S_{[i:\ell]}=\lambda$. 
\end{definition}
\begin{example}
   Let $\lambda={\textcolor{blue}{\mA\mA\mA\mC}}\in\Sigma^4$. The $\lambda$-labeling of the two sequences $S_1,S_2\in\Sigma^{20}$ is presented below. \vspace{-2ex}
\[\arraycolsep=1.4pt\def\arraystretch{2.2}
\begin{array}{cccccccccccccccccccccc}
        S_1 & = &\mA&\mA& \textcolor{blue}{\mA}&\textcolor{blue}{\mA}&\textcolor{blue}{\mA}&\textcolor{blue}{\mC}& \mT&\mG&\mT&\mG&\mC&\mA&\mT&\mA& \textcolor{blue}{\mA}&\textcolor{blue}{\mA}&\textcolor{blue}{\mA}&\textcolor{blue}{\mC}& \mC& \mG \vspace{-3ex}\\
        \lambda(S_1) & = &0&0& 1&1&1&1& 0&0&0&0&0&0&0&0& 1&1&1&1& 0&0
\end{array}
\]
\vspace{-4ex}
\[\arraycolsep=1.4pt\def\arraystretch{2.2}
\begin{array}{{cccccccccccccccccccccc}}
        S_2 & = &\textcolor{blue}{\mA}&\textcolor{blue}{\mA}&\textcolor{blue}{\mA}&\textcolor{blue}{\mC}& \textcolor{blue}{\mA}&\textcolor{blue}{\mA}&\textcolor{blue}{\mA}&\textcolor{blue}{\mC}& \mT&\mC&\mC&\mG&\mG&\mA&\mT&\mG&\mT&\mG&\mG&\mA \vspace{-3ex}\\
        \lambda(S_2) &= & 1&1&1&1&1&1&1&1&  0&0&0&0&0&0&0&0&0&0&0&0
\end{array}
\]
\end{example}
\vspace{-1ex}
In this work, we will be interested in labeling using multiple labels. Let $\Lambda=\{\lambda_1,\lambda_2,\ldots,\lambda_t\}$ be a set of $t$ labels $\lambda_i\in\Sigma^*$. The $\Lambda$-labeling function corresponds to labeling with all the $t$ labels in $\Lambda$ together and is defined as follows. 
\vspace{-0.5ex}
\begin{definition}
    Let $\Lambda = \{\lambda_1,\ldots,\lambda_t\}$ be a set of $t$ labels ${\lambda_i\in\Sigma^{\ell_i}}$.  The \emph{$\Lambda$-labeling function}, $\Lambda:\Sigma^*\to\{0,1\}^*$, is a mapping function that corresponds to the $\lambda$-labeling with all the labels $\lambda\in\Lambda$. The $\Lambda$-labeling of a sequence $S\in\Sigma^n$ is defined as the bitwise OR of the $t$~binary sequences ${\lambda_i}(S)$, $i\in[t]$, i.e.,
    \[
        \Lambda(S) \triangleq {\lambda_1}(S)\lor {\lambda_2}(S) \lor \cdots \lor {\lambda_t}(S),
    \]
    where $\lor$ stands for the bitwise OR operation.
\end{definition}

In other words, for a set of $t$ labels $\Lambda=\{\lambda_1,\ldots,\lambda_t\}$, the $\Lambda$-labeling of a sequence $S\in\Sigma^n$ is equal to the binary sequence $\Lambda(S)\in \{0,1\}^n$ in which $\Lambda(S)_{[i;\ell_j]}=(1\ 1\ \cdots\ 1)$ if and only if $S_{[i;\ell_j]}=\lambda_j\in \Lambda$ for some $j\in[t]$.

\vspace{-0.5ex}
\begin{example}
    Let $\Lambda=\{\textcolor{blue}{\mA\mA\mA\mC}, \textcolor{red}{\mC\mC}, \textcolor{teal}{\mG\mT\mG}\}$. 
    The $\Lambda$-labeling of the two sequences $S_1,S_2\in\Sigma^{20}$ is given as follow. 
    \vspace{-2ex}
\[\arraycolsep=1.4pt\def\arraystretch{2.2}
\begin{array}{cccccccccccccccccccccc}
        S_1 & = &\mA&\mA& \textcolor{blue}{\mA}&\textcolor{blue}{\mA}&\textcolor{blue}{\mA}&\textcolor{blue}{\mC}& \mT&\textcolor{teal}{\mG}&\textcolor{teal}{\mT}&\textcolor{teal}{\mG}&\mC&\mA&\mT&\mA& \textcolor{blue}{\mA}&\textcolor{blue}{\mA}&\textcolor{blue}{\mA}&\textcolor{Mulberry}{\mC}& \textcolor{red}{\mC}& \mG \vspace{-3ex}\\
        \Lambda(S_1) & = &0&0& 1&1&1&1& 0& 1&1&1& 0&0&0&0& 1&1&1&1& 1&0
\end{array}
\]
\vspace{-4ex}
\[\arraycolsep=1.4pt\def\arraystretch{2.2}
    \begin{array}{{cccccccccccccccccccccc}}
        S_2 & = &\textcolor{blue}{\mA}&\textcolor{blue}{\mA}&\textcolor{blue}{\mA}&\textcolor{blue}{\mC}& \textcolor{blue}{\mA}&\textcolor{blue}{\mA}&\textcolor{blue}{\mA}&\textcolor{blue}{\mC}& \mT&\textcolor{red}{\mC}&\textcolor{red}{\mC}&\mG&\mG&\mA&\mT&\textcolor{teal}{\mG}&\textcolor{teal}{\mT}&\textcolor{teal}{\mG}&\mG&\mA \vspace{-3ex}\\
        \Lambda(S_2) &= & 1&1&1&1&1&1&1&1&  0& 1&1& 0&0&0&0& 1&1&1& 0&0
    \end{array}
\]
\end{example}

\vspace{-1ex}
\begin{definition}\label{S-uniquely decodable}
    A \emph{labeling code} $\cC$ is a collection of sets of labels $\{\Lambda_i\}_{i=1}^{M}$ of sizes $\{t_i\}_{i=1}^{M}$ wherein each  $\Lambda_i\in\cC$ is called a \emph{codeset}. Given a sequence $S\in\Sigma^*$, we say that a labeling code $\cC$ is \emph{$S$-uniquely-decodable} if for any two distinct codesets $\Lambda_1,\Lambda_2\in \cC$, we have that $\Lambda_1(S)\ne\Lambda_2(S)$. 
\end{definition}

We note that \autoref{S-uniquely decodable} is equivalent to saying that a labeling code $\cC$ is $S$-uniquely-decodable if any codeset $\Lambda\in\cC$ can be uniquely recovered given $S$ and the $\Lambda$-labeling of $S$, $\Lambda(S)$. 

\begin{definition}
    Let $\Lambda_1,\Lambda_2$ be two sets  and let $S\in\Sigma^*$ be a sequence. We say that $\Lambda_1,\Lambda_2$ are \emph{$S$-equivalent} and denote $\Lambda_1\equiv_S\Lambda_2$ if the labeling of $S$ with $\Lambda_1$ is identical to the labeling of $S$ with $\Lambda_2$, i.e.,  $\Lambda_1(S)=\Lambda_2(S)$.
\end{definition}

\begin{example} 
    Consider the code $\cC = \{\Lambda_{1}, \Lambda_{2},\Lambda_{3}\}$, where
    $$
    \Lambda_{1} =  \{\mA\mC\},\ \ \  \Lambda_{2}=\{\mA\mC,\mG\mT\},\ \ \  \Lambda_{3}=\{\mA, \mC\}.
    $$
    It can be verified that $\cC$ is $S$-uniquely-decodable for 
    $S = \mA\mC\mG\mT\mA\mA\mA\mT$, by observing that ${{\Lambda_{1}}(S) = 11000000}$, ${{\Lambda_{2}}(S) = 11110000}$, ${{\Lambda_{3}}(S) = 11001110}$,
    are all different. On the other hand, $\cC$ is not $S'$-uniquely-decodable for the sequence
    $S' = \mA\mC\mG\mT\mA\mC\mG\mT$, since 
    $
    {\Lambda_{1}}(S') = 11001100 = {\Lambda_{3}}(S')
    $, i.e., $\Lambda_1$ and $\Lambda_3$ are $S'$-equivalent.
\end{example}

\subsection{Problems Statement}\label{subsec:Problem Statement}
In order to represent information using DNA labeling, we need a large codebook $\cC$ which is $S$-uniquely-decodable, for some \emph{reference sequence} $S\in\Sigma^n$. To this end, we need to consider both the design of such codes as well as the selection of the reference sequence~$S$. These  objectives are formally stated as follows. 
\begin{problem}\label{problem: code given S}
    Given a reference sequence $S\in\Sigma^n$, find the maximum size of a labeling code $\cC$ which is $S\text{-uniquely-decodable}$ and efficient encoder and decoder algorithms for an $S\text{-uniquely-decodable}$ code $\cC_S$ that achieves this maximum value. That is, we want to find the value 
        \[
        M(S)\triangleq \max\{\ |\cC|\ :\ \cC \text{ is $S$-uniquely-decodable}\},
        \]
        and efficient encoder and decoder for an $S\text{-uniquely-decodable}$ code $\cC_S$ such that $|\cC_S| = M(S)$.
\end{problem}
\begin{problem}\label{problem: executable labels}
    Given a collection of $T$ labels ${\cV = \{\lambda_1,\lambda_2,\ldots,\lambda_T\}}$, referred as the \emph{executable labels}, find the following.
    \begin{enumerate}
    \item Given a reference sequence $S\in\Sigma^n$, find the value 
    \[
     M(S,\cV)  \triangleq  \max\left\{ |\cC|\ \middle\vert \begin{array}{l}
    \text{$\cC$ is $S$-uniquely-decodable,} \\
    \ \ \ \ \ \ \ \ \ \ \cC\subseteq \cP(\cV)
  \end{array}\right\},
    \]
    where $\cP(\cV)$ is the power set of $\cV$ (i.e., for any $\Lambda\in\cC$ we have $\Lambda\subseteq \cV$). 
    \item Given a positive integer $n$ find the value $$M(n,\cV)\triangleq\max_{S\in\Sigma^n}\left\{ M(S,\cV)\right\},$$ a reference sequence $S_{(n,\cV)}\in\Sigma^n$ such that 
    \[
    M(S_{(n,\cV)},\cV) = M(n,\cV),
    \]
    and an $S_{(n,\cV)}$-uniquely decodable code $\cC_{(n,\cV)}\subseteq\cP(\cV)$ with efficient encoder and decoder algorithms. 
    \end{enumerate}
\end{problem}

\autoref{problem: code given S} represents a theoretical question and we will describe some related results. \autoref{problem: executable labels} represents a variation of the model that encompasses the case of CRISPR labels~\cite{jinek2012programmable, chen2018efficient, cong2013multiplex, amit2021crispector} where $\lambda_i$ will be the corresponding potential edit sites.

\subsection{Basic Results using Periodicity}
\label{sec: problem 1}
We start with the simplest case of \autoref{problem: code given S} in which the reference sequence $S$ consists of a single run. That is, $S=\sigma^n$ for some  $\sigma\in\Sigma$ and positive integer $n$. For this case, we fully solve \autoref{problem: code given S} in the next lemma. 

\begin{lemma}\label{lem: S for minimal code}
    If $S$ is a sequence with a single run of the symbol $\sigma$, then for any $S$-uniquely-decodable labeling code $\cC$ we have that $M(S)= 2$. Furthermore,  the code $\cC_\sigma=\{\varnothing, \{\sigma\}\}$, is  $S$-uniquely-decodable. 
\end{lemma} 

\begin{IEEEproof}
To see that $\cC_\sigma$ is $S$-uniquely-decodable note that for $\Lambda_1=\varnothing$ and $\Lambda_2=\{\sigma\}$ we have that $\Lambda_1(S)$ is the \emph{all-zero} word while $\Lambda_2(S)$ is the \emph{all-one} word.

Let $\cC$ be an $S$-uniquely-decodable labeling code with maximum size. Note that for any label $\lambda\in\Sigma^*$, if $\lambda\ne \sigma^i$, for any integer $0< i\le |S|$, then $\lambda$ is not a substring of $S$ and $\lambda(S)$ is the \emph{all-zero} word. Otherwise, we have that $\lambda(S)$ is the \emph{all-one} word. By definition of the bitwise OR operation, for any codeset $\Lambda\in \cC$ we have that either $\Lambda\equiv_S \varnothing$  or $\Lambda\equiv_S\{\sigma\}$. Since $\cC$ is  $S$-uniquely-decodable, the latter implies that $|\cC|\le 2$.
\end{IEEEproof}

To extend \autoref{lem: S for minimal code} to more involved cases, we first present the definitions of a \emph{period}.

\begin{definition}
    For $S \in \Sigma^n$, $1 \leq \pi \leq n-1$ is called a \emph{period} of $S$ if $\pi|n$ and for all $1 \leq i \leq n-\pi+1$, ${S_{i} = S_{i+\pi}}$. Additionally, we let $\pi(S)$ be the minimal period of the sequence $S$ if such a period exists and otherwise $\pi(S)=n$.
\end{definition}

\begin{lemma} \label{lem: period two}
For any sequence $S$ with period $\pi(S) = 2$, it holds that $M(S) = 7$.
\end{lemma}
\begin{IEEEproof}
    Assume w.l.o.g. that $S=\mA\mC\mA\mC\ldots \mA\mC$.
     The only labels that can be considered are the ones that are subsequences of $S$, i.e., the possible labels are
     \vspace{-1.1ex}
     \[
     \bigcup_{t=0 }^{\frac{n}{2}-1}
     \left\{ (\mA\mC)^{t+1}, (\mC\mA)^{t+1}, (\mA\mC)^t\mA, (\mC\mA)^t\mC
     \right\} 
         \vspace{-1.1ex}
     \]
     
     Furthermore, for any $1\le t\le {n}/{2}$ we have that 
         ${\{\mA\mC\}\equiv_S \{(\mA\mC)^t\}}$, and
         ${\{\mC\mA\}\equiv_S \{(\mC\mA)^t\}}$,
     and for any ${1\le t\le {n}/{2}-1}$ we have that 
         $\{\mA\mC\mA\}\equiv_S \{(\mA\mC)^t\mA\}$, and
         ${\{\mC\mA\mC\}\equiv_S \{(\mC\mA)^t\mC\}}$.
     Thus, it is sufficient to consider only the labels in $\{\mA, \mC, \mA\mC, \mC\mA, \mA\mC\mA, \mC\mA\mC\}$. It can be verified that the labeling of $S$ by each of these labels is unique, and by considering the empty codeset we have an $S$-uniquely decodable code of size $7$. To see that no larger code is $S$-uniquely decodable, note that any possible codeset is $S \text{-equivalent}$ to some codeset that is composed only of labels in  $\{\emptyset, \mA, \mC, \mA\mC, \mC\mA, \mA\mC\mA, \mC\mA\mC\}$ and that any codeset that is a subset of the last six labels, is $S$-equivalent to a codeset that consists of a single label out of the latter six labels.
\end{IEEEproof}

The upper bound in \autoref{lem: period two} can be extended to any period $\pi$ by similar arguments. However, for $\pi>2$ this upper bound is not necessarily tight. This result is summarized in the following lemma while the proof is left for the full version of the paper.

\begin{lemma}
    For any sequence $S$, we have that 
    $$M(S)\le 2^{2\pi(S)-2} + 2^{\pi(S)} -1.$$
\end{lemma}

\section{Fixed-Length Labels}\label{sec: fix length}
In this section, we consider the special case in which all the labels have the same length. That is, we consider \autoref{problem: executable labels} in the special case where the executable labels are all the labels of length $\ell$, for some integer $\ell$, i.e., $\cV=\Sigma^\ell$. To this end, for a reference sequence $S$ and an integer $n$ we define $M_\ell(S)\triangleq M(S,\Sigma^\ell)$ and $M_\ell(n) \triangleq M(n,\Sigma^\ell)$. Similarly, we use the notations $S_{(n,\ell)}$ and $\cC_{(n,\ell)}$ for the reference sequence and code defined in \autoref{problem: executable labels}.2. 

\begin{definition}
    A labeling code $\cC$ is called an \emph{$\ell$-labeling code} if for any codeset $\Lambda\in\cC$ and any $\lambda\in\Lambda$ it holds that $|\lambda|=\ell$.
\end{definition} 

\vspace{-0.5ex}
We start with an upper bound on $M_\ell(n)$ for any two integers $0<\ell\le n$. Then, we show that in the special case where $\ell\ge\log_4(n-\ell+1)$, this bound is tight. Before we present the bound and an explicit construction that meets it for $\ell\ge\log_4(n-\ell+1)$, let us define $\eta(n,\ell)$ to be the number of binary sequences of length $n$ in which each run of consecutive \emph{ones} is of length at least $\ell$. Additionally, define the constraint $\cT_\ell$ to be the set of all binary sequences in which any run of \emph{ones} is of length at least~$\ell$. This constraint is strongly related to the 
well-known \emph{run-length limited} (\emph{RLL}) constraint, which is described in the next definition~\cite{marcus2001introduction}.

\vspace{-0.5ex}
\begin{definition}
A binary sequence satisfies the \emph{$(d,k)$-RLL constraint} if between every two consecutive \emph{ones}, there are at least $d$ \emph{zeros} and there is no run of \emph{zeros} of length $k+1$. Denote the set of all sequences of length $n$ that satisfy the $(d,k)$-RLL constraint by $\cC_{d,k}(n)$. 
\end{definition}

\vspace{-0.5ex} We note that the constraint $\cT_\ell$ is equivalent to the $(\ell,\infty)$-RLL constraint (replacing the roles of zeros and ones). It has been proven that for any constant $\ell$, ${\mathsf{cap}(\cC_{\ell,\infty})= \log_2\rho}$ where $\rho$ is the largest real root of $x^{\ell+1}-x^{\ell}-1$~\cite{McLaughlin95}, which implies the following corollary.

\begin{corollary}
\label{cor: RLL capacity}
Let $\ell$ be a positive integer. We have that ${M_\ell(n) \le 2^{\Theta(n\cdot\log_2\rho)}}$, where $\rho$ is the largest real root of $x^{\ell+1}-x^{\ell}-1$. Furthermore, for a constant $\ell$, it holds that  $\lim_{n\rightarrow \infty}\frac{\log(M_\ell(n))}{n} \leq \log_2\rho$.
\end{corollary} 

\vspace{-0.5ex}
\begin{theorem}\label{th: upper bound}
Let $\ell$ and $n$ be positive integers. For any reference sequence $S\in\Sigma^n$, we have that $M_\ell(S)\le  \eta(n,\ell)$, and hence $M_\ell(n)\le \eta(n,\ell)$.
\end{theorem}
\begin{IEEEproof}
    Let $\cC$ be an $\ell$-labeling code. Note that by the definition of $\ell$-labeling code, for any $\Lambda \in \cC$, we have that $\Lambda(S)$ is a binary string of length $|S|$, in which any run of \emph{ones} is of length $\ell$ or more. Additionally, since $\cC$ is $S$-uniquely-decodable, for any  $\Lambda_1, \Lambda_2 \in \cC$, we have that $\Lambda_1(S)\ne \Lambda_2(S)$, and thus,  $|\cC|\le \eta(|S|,\ell)$.
\end{IEEEproof}

Next, we show that the bound in \autoref{th: upper bound} is tight by presenting an explicit construction for a code $\cC_{(n,\ell)}$ that meets this bound with equality when $S$ is an \emph{$\ell$-repeat-free} sequence, i.e., any substring of length $\ell$ in $S$ is unique~\cite{elishco2021repeat}. Note that such a sequence $S$ corresponds to a trail in the de Bruijn graph. 

\begin{construction}\label{const: meet uppper bound}
    Let $S$ be an $\ell$-repeat-free sequence of length~$n$ over $\Sigma$ and let $\cT_\ell(n)\triangleq \cT_\ell\cap\{0,1\}^n$ be the set of all binary strings of length $n$, in which any run of \emph{ones} is of length $\ell$ or more. For any $X=(x_1,\ldots, x_n)\in \cT_\ell(n)$, we define the codeset $\Lambda_X$ as follows. 
    \begin{enumerate}
        \item Initialize an empty set $\Lambda_X =\varnothing$. 
        \item For any run of \emph{ones} in $X$ of length $\ell'\geq \ell$, $X_{[i;\ell']}$:
        \begin{enumerate}
            \item Add the labels
            $S_{[i;\ell]}, S_{[i+\ell;\ell]}, \ldots  , S_{[i+\ell\cdot \lfloor\ell'/\ell\rfloor;\ell]}$ into $\Lambda_X$.
            \item  If  $\ell'/\ell$ is not an integer, add the label $S_{[i+\ell'-\ell;\ell]}$ into $\Lambda_X$.
        \end{enumerate}
    \end{enumerate}  
    Finally, we define  
    $\cC_{(n,\ell)} = \{\Lambda_X: X\in \cT_\ell(n)\}$.
\end{construction}

\begin{example}
Let $S\in\Sigma^{25}$, and $X\in\cT_4(25)$ be the following sequences, 
\vspace{-2ex}
\[
\arraycolsep=0.8pt\def\arraystretch{2.0}
\begin{array}{{ccccccccccccccccccccccccccc}}
S & = & \mA& \mA& \mA& \mA& \mC& \mA& \mA& \mA& \mG& \mA& \mA& \mA& \mT& \mA& \mA& \mC& \mC& \mA& \mA& \mC& \mG& \mA& \mA& \mC& \mT, \vspace{-1.5ex}\\
X & = & 1& 1& 1& 1& 1& 1& 1& 1& 1& 1& 1& 0& 0& 0& 0& 0& 1& 1& 1& 1& 0& 0& 0& 0& 0.
\end{array}
\vspace{-1ex}
\]
The sequence $S$ is a $4$-repeat-free sequence, and the codeset $\Lambda_X$ which is obtained by \autoref{const: meet uppper bound} for the binary sequence $X$ is
$\Lambda_X = \left\{ 
\mA\mA\mA\mA, \ \mC\mA\mA\mA, \ \mA\mG\mA\mA, \ \mC\mA\mA\mC
\right\}$,
where the first three labels correspond to the first run of \emph{ones} (in $X$) and the last label corresponds to the second run of \emph{ones}. 
\end{example}
\begin{theorem}
    For any $\ell$-repeat-free  sequence $S\in\Sigma^n$, the code $\cC_{(n,\ell)}$ obtained by \autoref{const: meet uppper bound} is an $S$-uniquely-decodable $\ell$-labeling code of size $\eta(n,\ell)$. 
\end{theorem}
\begin{IEEEproof}
     For any $1\le i\le n-\ell+1$ the subsequence $S_{[i;\ell]}$ is of length $\ell$ by definition. Hence, for any $X$, $\Lambda_X$ that is defined by the algorithm can only contain labels of length exactly $\ell$. That is, $\cC_{(n,\ell)}$ is an $\ell$-labeling code. Furthermore, for any $X\in\cT_\ell(n)$, we have that any label $\lambda\in\Lambda_X$ appears only once as a subsequence of $S$ and hence $\Lambda_X(S)=X$. Thus, for any two distinct sequences $X,Y\in \cT_\ell(n)$,
     we have that $\Lambda_X(S)=X\ne Y=\Lambda_Y(S)$ which implies that $\cC_{(n,\ell)}$ is an $S$-uniquely-decodable code of size $\eta(|S|,\ell)$. 
\end{IEEEproof}

We note that to encode a binary message, we first need to encode it into a sequence $X\in\cT_\ell(n)$ and then apply \autoref{const: meet uppper bound} on $X$. An efficient encoder of the $\cT_\ell$ constraint can be achieved by constructing a deterministic finite automaton that accepts all the constrained words, and then utilizing the state-splitting algorithm \cite{adler1983algorithms} to design encoders with rate approaching
the capacity.  

The following lemma shows that the size of $\ell$-labeling codes is maximized for $S$ which is $\ell$-repeat-free (if such $S$ exists). 
\begin{lemma}\label{lem: not repeat-free}
    If $S\in\Sigma^n$ is not $\ell$-repeat-free, then $$M_\ell(S)<\eta(n,\ell).$$ 
\end{lemma}

The proof of \autoref{lem: not repeat-free} is based on the fact that for any $S$ which is not $\ell$-repeat-free there exists a sequence in $\cT_\ell(|S|)$ which can not be obtained as a labeling of~$S$. The detailed proof of \autoref{lem: not repeat-free}  is left for the full version of the paper.

Let $n$ and $\ell$ be two positive integers and let $S\in\Sigma^n$. \autoref{th: upper bound} and \autoref{lem: not repeat-free} state that  if there exists an $\ell$-repeat-free sequence of length $n$, then an $S$-uniquely-decodable $\ell$-labeling code $\cC$ is of maximum size if and only if $S$ is $\ell$-repeat-free. However, it should be noted that for any sequence $S$ of length $n$, if $S$ is $\ell$-repeat-free then $|S|\le 4^\ell+\ell-1$. That is, \autoref{const: meet uppper bound} is relevant only for $\ell\ge \log_4(|S|-\ell+1)$. 

\begin{corollary}
    If $\ell< \log_4(n-\ell+1)$ then $M_\ell(n)<\eta(n,\ell).$
\end{corollary}

\begin{corollary}
    Let $S\in\Sigma^n$ be a reference sequence and let ${\ell\geq \log_4(n-\ell+1)}$. We have that $M_\ell(S) = M_\ell(n)$ if and only if $S$ is an $\ell$-repeat-free sequence. 
\end{corollary}

    \vspace{-1.9ex}
\begin{corollary}
$\ $
$$M_\ell(n) = \eta(n,\ell) \text{ if and only if } \ell\geq \log_4(n-\ell+1).$$
\end{corollary}

In the following lemma and corollary the value $\eta(n,\ell)$ is analyzed.

\begin{lemma}\label{lem: eta} For any two positive integers $n,\ell$, it holds that 
\[
\eta(n,\ell) = \sum_{t=0}^{\lfloor \frac{n+1}{\ell+1}\rfloor}\binom{n-t(\ell-1)+1}{2t}.
\]
\end{lemma}

\begin{IEEEproof}
    Let $A$ be the set of binary strings of length $n$ in which each run of \emph{ones} is of length at least $\ell$. We start by partitioning $A$ into two disjoint sets,
    \begin{align*}
        A^0& =\left\{(x_1,\ldots,x_n)\in A:\ x_1=0\right\}\\
        A^1& =\left\{(x_1,\ldots,x_n)\in A:\ x_1=1\right\}.
    \end{align*}
    Clearly, $\eta(n,\ell) = |A| = |A^0\cup A^1| = |A^0| + |A^1|$ since $A^0\cap A^1=\emptyset$. Next, let us calculate $|A^0|$, the analysis for $|A^1|$ is done similarly. We have that
    $|A^0| =  \sum_{r=1}^n  |A^0_{r}|$,
    where $A^0_{r}$ is the set of all sequences in $A^0$ that have exactly $r$ runs (of zeros/ones). Since the first run of any sequence in $A^0$ is a run of \emph{zeros}, any sequence in $A^0_{r}$ contains $r_1 = \lceil(r-1)/2\rceil$ runs of \emph{ones} and $r_0=r-r_1 =\lfloor(r+1)/2\rfloor $ runs of \emph{zeros}.
    Additionally, each run of \emph{zeros} must contain at least a single \emph{zero} and each run of \emph{ones} must contain at least $\ell$ \emph{ones}. Let $\CC{m}{b}$ denote the number of $b$-element combinations of $m$ objects, with repetitions. We have that  
    \begin{align*}
        |A_{r}^0| & = \hspace{-2pt}\CC{r}{n-\ell r_1-r_0} \hspace{-2pt}= \hspace{-2pt}\CC{r}{n-\lceil(r-1)/2\rceil(\ell-1) -r } \\
        & = \hspace{-2pt}\binom{n-\lceil(r-1)/2\rceil(\ell-1)- 1}{r-1}
    \end{align*}
    and 
        \vspace{-2.5ex}
    \begin{align*}
        |A^0| =  \sum_{r=1}^n  \binom{n-\lceil(r-1)/2\rceil(\ell-1)- 1}{r-1}.
    \end{align*}
Similarly, we have that 
\begin{align*}
    |A^1| & = \sum_{r=1}^n \CC{r}{n-\lfloor(r+1)/2\rfloor(\ell-1)-r}\\
    & = \sum_{r=1}^n \binom{n-\lfloor(r+1)/2\rfloor(\ell-1)-1}{r-1}
\end{align*}
and thus 
\vspace{-1.5ex}
\begin{align*}
    \eta(n,\ell) = & \sum_{r=1}^n
    \binom{n-\lceil\frac{r-1}{2}\rceil(\ell-1)- 1}{r-1} \\ & +\sum_{r=1}^n \binom{n-\lfloor\frac{r+1}{2}\rfloor(\ell-1)-1}{r-1}.
\end{align*}
By utilizing the identity $\binom{n-1}{k-1}+\binom{n-1}{k}=\binom{n}{k}$, and by considering even and odd values of $n$ separately, it can be shown that the latter expression is equal to
\[
\eta(n,\ell) =  \sum_{t=0}^{\lfloor \frac{n+1}{\ell+1}\rfloor}\binom{n-t(\ell-1)+1}{2t}.
\vspace{-1.5ex}\]
\end{IEEEproof}

 Recall that the upper bound in \autoref{cor: RLL capacity} is evaluated under the assumption that $\ell$ is a constant. Since Construction~\ref{const: meet uppper bound} requires that the value of $\ell$ will grow as a function of $n$, the latter capacity does not represent the achievable rates of our construction. Hence, in the next theorem, we present the asymptotic behavior of $\eta(n,\ell)$ for the case $\ell = c \log_4(n)$, for $c\ge 1$. The proof follows from analyzing the expression given in \autoref{lem: eta} for $\ell = c \log_4(n)$ when $n\to \infty$.

\begin{theorem}
    Let $n$ be a positive integer and let $\ell = c\log_4(n)$ for $c\ge 1$. Additionally, let $S$ be an $\ell$-repeat-free sequence and let $\cC_{(n,\ell)}$ be the code obtained by \autoref{const: meet uppper bound}. It holds that
    \[
    M_\ell(n)=M_\ell(S)=|\cC_{(n,\ell)}| = \eta(n,\ell) = 2^{\Theta\left(\frac{\log\log(n)}{\log(n)}\cdot n\right)}.
    \]
\end{theorem}

To conclude this section, we discuss the case where we are interested in an $\ell$-labeling code which is $S$-uniquely-decodable, while the sequence $S$ is not $\ell$-repeat free. We start by proving a sufficient condition for an $\ell$-labeling code to be $S$-uniquely-decodable for a given sequence $S$ (not necessarily $\ell$-repeat-free). 
To this end, for an $\ell$-labeling code $\cC$ we define $\cC_{(S)} \triangleq \{\Lambda\cap W_\ell(S):\ \Lambda\in\cC\}$.


\begin{lemma}\label{lem:C_l sufficient cond basic}
    Let $0<\ell\le n$ be two integers and let $S\in\Sigma^n$. Additionally let $\cC$ be an $\ell$-labeling code. Then, $\cC$ is $S$-uniquely-decodable if the following two conditions hold:
    \begin{enumerate}
        \item $|\cC|=|\cC_{(S)}|$. 
        \item For any $\Lambda\in\cC$ we have that no proper prefix of a label $\lambda\in\Lambda$ is a proper suffix of a label $\lambda'\in\Lambda$.
    \end{enumerate}
\end{lemma}

The first condition in \autoref{lem:C_l sufficient cond basic} implies that all the codesets in $\cC$ are different when ignoring labels that do not appear as substrings of $S$. The second condition is that any codeset in $\cC$ is a \emph{non-overlapping code} \cite{levy2018mutually, levenshtein1970maximum, yazdi2018mutually, blackburn2015non}, that is, for any codeset  $\Lambda\in\cC$ we have that no proper prefix of a label $\lambda\in\Lambda$ is a suffix of a  label $\lambda'\in\Lambda$. In fact, this condition is too strong and we only need the latter to hold concerning the \emph{embeddings} of the labels in $S$. Hence, we prove the following weaker sufficient condition instead.  

\begin{theorem}\label{th: C_l sufficient cond non-overlap in S}
    Let $S\in\Sigma^n$ and let $\cC$ be an $\ell$-labeling code. Then, $\cC$ is $S$-uniquely-decodable if the following two conditions hold:
    \begin{enumerate}
        \item $|\cC| = |\cC_{(S)}|$.
        \item For any $\Lambda\in \cC_{(S)}$ and any $\lambda,\lambda'\in\Lambda$, if there exists a sequence $\bfw\in\Sigma^*$ such that $\lambda$ is a prefix of $\bfw$, $\lambda'$ is a suffix of $\bfw$, and $|\bfw|<2\ell$, then $\bfw$ is not a substring of $S$. 
    \end{enumerate}
\end{theorem}

\begin{IEEEproof}
    Let $\cC$ be an $\ell$-labeling code that satisfies the two conditions in the claim. 
    We first note that $|\cC|=|\cC_{(S)}|$ implies that for any two codesets in $\cC$, their subsets which contain only the $\ell$-substrings of $S$ are distinct. Hence we can ignore the labels that are not substrings of $S$ and prove that the second condition     guarantees that for any $\Lambda\in\cC$, the codeset $\Lambda$ can be uniquely determined from $\Lambda(S)$ and $S$. 
    
    For any $\Lambda\in\cC$, the second condition implies that the sequence $\Lambda(S)$ is a binary sequence of length $n$ in which any run of \emph{ones} is of length~$j\ell$ for some integer $j\ge 1$. Furthermore, for any run of \emph{ones} in $\Lambda(S)$ there is a unique way to partition it to non-overlapping segments of length $\ell$. Thus, for any run of $j\ell$ \emph{ones} in $\Lambda(S)$, $j\ge 1$, we have a unique set of non-overlapping labels of length $\ell$ that can create this run. That is, $\Lambda$ can be uniquely determined given $S$ and $\Lambda(S)$, which completes the proof. 
\end{IEEEproof}

Assume $\ell<\log_4(n-\ell+1)$, using \autoref{lem:C_l sufficient cond basic}, we can construct an $\ell$-labeling code which is $S$-uniquely decodable for a reference sequence $S\in\Sigma^n$ as follows. 

\begin{construction} \label{cons: not rf}
Given $\ell<\log_4(n-\ell+1)$ and a sequence $S\in\Sigma^n$, construct an $\ell$-labeling code $\cC$ as follows.
\begin{enumerate}
    \item  Let $\cN_\ell$ be the non-overlapping code of length $\ell$ over $\Sigma$ given in \cite{blackburn2015non}. 
    \item Let $\cN_{(\ell,S)}\triangleq \cN_\ell\cap W_\ell(S)$.
    \item  Let $\cC\triangleq \cP(\cN_{(\ell,S)})$, where $ \cP(\cN_{(\ell,S)})$ is the power set of $\cN_{(\ell,S)}$.
\end{enumerate}
\end{construction}

It was proven in \cite{blackburn2015non}, that 
$|\cN_\ell| \ge 63\cdot \frac{4^{\ell-5}}{\ell}$,
which implies the following lower bound on $M_\ell(S)$ and $M_\ell(n)$.
\begin{claim}\label{cla: non overlapping}
    For any $S\in\Sigma^n$ we have that $M_\ell(S)\ge 2^{|\cN_{(\ell,S)}|}$. In particular, 
    $M_\ell(n)\ge 2^{63\cdot \frac{4^{\ell-5}}{\ell}}$.
\end{claim}
\begin{IEEEproof}
    Let $S$ be a sequence of length $n$ such that the prefix of length $4^\ell+\ell-1$ of $S$ is a de Bruijn sequence. In this case, $\cN_{(\ell,S)} = \cN_\ell$ and hence the size of the code $\cC$ from \autoref{cons: not rf} is 
    $|\cC| = 2^{|\cN_{(\ell,S)}|}\ge 2^{63\cdot \frac{4^{\ell-5}}{\ell}}.$
    Hence, $M_\ell(S)\ge 2^{63\cdot \frac{4^{\ell-5}}{\ell}}$ which implies the claim of the theorem. 
\end{IEEEproof}

While \autoref{cons: not rf} and the bound in  \autoref{cla: non overlapping} can be used for any sequence $S$, in case we can use any reference sequence of length $n$, these results can be improved, as shown in the next theorem.
\begin{theorem}
    For integers $n$ and $\ell$, if $\ell<\log_4(n-\ell+1)$ then
    $$M_\ell(n)\ge \eta(4^\ell+\ell-1, \ell) = 2^{\Theta\left(\frac{\log(\ell)}{\ell}\cdot (4^\ell+\ell-1)\right)}$$
\end{theorem}
\begin{IEEEproof}
    To see that the claim in the theorem holds, let $S$ be a sequence of length $n$ such that the prefix of length $4^\ell+\ell-1$ of $S$ is a de Bruijn sequence and  the code $\cC_{4^\ell+\ell-1,\ell}$ obtained by \autoref{const: meet uppper bound} where the decoding is done using the  de Bruijn prefix of $S$. 
\end{IEEEproof}

\section{Conclusions}
This study presents a theoretical framework for utilizing DNA molecules to encode information, intending to circumvent expensive DNA synthesis and facilitate a more practicable approach to certain applications. Our proposed methodology suggests to leverage established biochemical techniques commonly employed in medical and biological research, such as CRISPR-Cas9 and gRNA reagents for labeling. Through comprehensive exploration, we establish upper bounds on achievable codes under specific conditions and introduce an efficient encoder-decoder pair optimized for maximal code size. 

To make further progress, future research should prioritize adapting the model, constructions, and bounds to a more realistic scenario that incorporates noise. One form of noise to consider is the potential for a small number of labels to be missing. This can result from labeling failures at specific sites or false negatives during the identification of labeled locations. Additionally, it’s essential to address synchronization errors that may arise when determining labels’ positions.

\newpage
\balance
\bibliographystyle{IEEEtran}
\bibliography{arxib}
\clearpage
\end{document}